# Managing Expert Disagreement for the Policy Process and Beyond

Ulrike Hahn[1], Jens Koed Madsen[2] & Chris Reed[3]

u.hahn@bbk.ac.uk
*Birkbeck, University of London, U.K.[1]*
*London School of Economics, U.K.[2]*
*University of Dundee, U.K.[3]*

**Abstract.** In this paper, we outline a new proposal for communicating scientific debate to policymakers and other stakeholders in circumstances where there is substantial disagreement within the scientific literature. In those circumstances, it seems important to provide policy makers both with a useful, balanced summary that is representative of opinion in the field large, and to transparently communicate the actual evidence-base. To this end, we propose the compilation of argument maps through a collective intelligence process; these maps are then given to a wide sample of the relevant research community for evaluation and summary opinion in an IGM style IGM style poll (see igmchicago.org), which provides a representative view of opinion on the issue at stake within the wider scientific community. Policymakers then receive these two artefacts (map and poll) as their expert advice. Such a process would help overcome the resource limitations of the traditional expert advice process, while also providing greater balance by drawing on the expertise of researchers beyond the leading proponents of particular theories within a field. And, the actual evidence base would be transparent. In this paper, we present a pilot project stepping through the map building component of such a policy advice scheme. We detail process, products, and issues encountered by implementing in the OVA (Online Visualisation of Argument tool, ova.arg-tech.org) an argument map with sample evidence from the behavioural literature on communicating probabilities, as a central issue within pandemic.

**Keywords.** Argument mapping; policy process; evidence communication;

## 1. Introduction

The pandemic has brought the interface between science and policy to the forefront like few other events. Policymakers were forced to make decisions of enormous consequence on the basis of scientific evidence that spanned many disciplines and was itself in many parts rapidly developing. It was thus clear early on that the crisis would place enormous strain on extant scientific advisory systems. The standard process for such systems is arguably resource intensive, prone to bias and lack of representative breadth, and frequently non-transparent to the wider scientific community as well as the general public. These features become particularly problematic in a situation such as the pandemic when a broad range of expertise and perspective is required in order to make sense of a complex, yet partial and continuously evolving set of scientific considerations that are of huge consequence to society. In this situation, it seems desirable to search for novel approaches that might harness *collective intelligence* in order to generate rapid, robust and transparent scientific communications on which policy makers might base their actions.

In this paper, we present a novel suggestion for harnessing the full resources of a scientific community in order to generate artifacts that, as accurately and evenhandedly as possible, portray a current body of evidence. In particular, we suggest that an extant literature may be compiled into an argument map, thus providing a transparent reflection of the evidence base that would provide the wider scientific community, policymakers, and stakeholders, with the basis for and even handed an assessment of the evidence. Such a map could be given to a wider community of researchers for evaluation, and together with that evaluation provide a summary report that indicates clearly and transparently what disagreement and uncertainty exist, while still providing overall, summary recommendations where possible and appropriate.

An initial sketch of the general process we propose runs as follows: 1) identify relevant literature; 2) mark up individual arguments within the debate; 3) transfer those arguments to a machine readable template; 4) build argument map; 5) invite proponents of competing theories for comment, clarification, or additional inputs; 6) invite members of the wider scientific community of relevance to "vote" and comment; 7) aggregate community response in an IGM forum style; 8) publish map + response.

In the following, we elaborate on the background to this proposal by describing in more detail the theoretical motivations, available tools, and possible implementations. We then describe an initial case study involving a limited and partial implementation of the map building process (Steps 1-4 above). Despite being limited in scope, it serves to highlight key conceptual and technical issues that need to be addressed in order to run a full-scale version of this process. We then contrast such an approach with other extant approaches to evidence aggregation. We conclude with suggestions for future research and development.

## 2. Background

In the standard policy process, advisors sound out experts in a field, identify issues, and identify individuals who are then called upon to help compile reports on policy relevant scientific issues, whether as part of an ad hoc body, or a semi permanent standing committee. This process is time and labour intensive, and advice is given in addition to normal responsibilities and duties, and typically without remuneration.

Recognised scientific experts are a scarce resource; recognised experts that are able and willing to interact with the policy process are an even scarcer resource. Yet the wider scientific community, ranging from post graduate students through early career researchers to senior academics, constitutes a wide pool of expertise (of varying levels) that typically remains largely untapped for the purposes of policy advice. It would thus be desirable to utilise these significant untapped resources, particularly in times of crisis.

Collective intelligence (see Suran et al. 2020) provides a potential solution for dealing with high stakes resource intensive problems, by drawing on wider communities. Collective intelligence might be defined as the capacity to solve complex problems via a group or collective, typically aided by technological tools.

But there are further, entirely epistemic, reasons why devolving scientific advice to an entire community might be desirable. This stems from the fact that science is intrinsically open ended, incomplete, and subject to change. This means that theoretical differences among scientists are, if anything, the norm, not the exception. This has implications for science communication, and more so the more high stakes the decisions that are ultimately to be taken based on those scientific assessments.

Wherever possible scientific experts should communicate a scientific consensus, glossing over theoretical differences that do not make different predictions in terms of the policy issue in question. But how should the case of real, policy relevant, scientific disagreement be dealt with?

In these circumstances, relying on select, high profile, experts may be less than ideal for a number of reasons: 1) leading experts, even where well intentioned, may struggle to reconstruct ‘even handed’ descriptions of debates they have been involved in; 2) scientific debates may will involve “deep” disagreements that include divergent opinions on what counts as relevant data to be explained by a theory (scope), the quality of data, and desiderata for theories themselves (simplicity, explanatory goodness, precision, etc.); 3) scientific debates are ultimately resolved not by the proponents of theories themselves but the wider scientific community.

The present paper seeks to sketch out a strategy/tool for dealing with expert disagreement that respects these 3 constraints by i) replacing individual expert views with a community built map of the evidence that is available for public scrutiny, which ii) seeks only to explicate/visualise the argumentative discourse without evaluation and iii) uses transparent aggregation of verdicts from the wider scientific community to establish ‘recommendations’ in a way that communicates the range of opinion and allows policy makers to form their own evaluation.

In short, we propose the creation of an argument map of the arguments/evidence underlying the opposing positions, combined with a transparent community “evaluation and aggregation”. The deliverables to policy makers are the artefacts this process creates.

Our main focus in this paper will be on the creation of the argument map, but we start with some thoughts on the evaluation and aggregation of such a map through a wider research community.

As it seems unlikely that all key proponents in a theoretical debate within a scientific community would be involved directly in the construction of a map, a first evaluation step might see the argument map handed to them for comment. In this step, those experts directly involved in the debate within the primary research literature could signal their agreement with the basic construal. Or they could ask for revisions. Ideally, the argument map would thus receive approval from the key researchers involved. For reasons just outlined above, however, one might want to receive a final assessment on the status of the debate from a wider community of researchers within that field, beyond just those authors whose published work has been included within the map. Gauging the perspective of that wider community of relevant scholars seems the most robust way to gauge the level of consensus and agreement on what the evidence currently shows.

One way to do this, that is transparent and accurately reflects agreement, is a poll in the style of the Chicago Booth Initiative on Global Markets (see igmchicago.org). These polls pose questions to a panel of economic experts, asking them not just to rate the polls’ preprepared statement but also providing an opportunity to comment or provide additional resources. Results are then communicated to the public both an aggregate and with an opportunity to look at individual opinions. This, we think strikes a nice balance between both providing an opportunity to determine the overall distribution of opinions and still probe the views of given individuals.

In the current context, one could implement voting and commenting in different ways. It would be desirable, however, to have evaluators provide both a summary assessment of the confidence they have in the accuracy of each of the competing theories, and provide judgements of the strength of individual arguments or pieces of evidence. In addition to this, they could provide free-form comments in case there are elements of the maps that have not been included, but that should be included.

Closer consideration would obviously have to be given to the selection of the researchers to be consulted in such a poll. There are clearly many ways to solicit those individuals, and to determine qualifying criteria (for one possibility see, Lewandowsky et al., 2021). But our vision is that this would include a broad selection of researchers in the field, not just leading exponents.

Combined with the fact that evaluation of an argument map summarising evidence in a domain should require considerably less effort then producing a full, written, report of such evidence, this should enable much broader participation and representativeness than is typically seen in the policy process. This matters because epistemic diversity has direct implications for the accuracy of collective judgements (see e.g., Page, 2006; Hahn, 2022).

Clearly, there remain many details of such a process to be worked out, tried and tested empirically. In the remainder of this paper, however, we now focus on the construction of the argument map itself.

### *2.1. Argument mapping*

The idea that complex debate or evidential relationships might be visualised graphically through some type of map has a long tradition. An early example is Wigmore's Chart Method (Wigmore 1913). The idea here was to use a chart including well-defined types of arcs to represent different types of evidence, facts and relations between them in order to summarise the relevant details of legal cases. A further milestone in the history of argument mapping is Toulmin's (1958) monograph The Uses of Argument. In this work, Toulmin sought to develop a framework for every day informal argument, given the fact that logical (deductive) reasoning has little to say about such argument. As part of this, he proposed a representational framework for laying out the structure of arguments that has continued to enjoy popularity to this day. His scheme distinguishes between fact or observations (a datum) which might be put forward in favour of a claim. Linking datum, and claim is the so-called warrant, which itself may have a further backing.. Still used widely in the context of critical thinking pedagogy it has inspired many subsequent schemes. More recently, Toulmin's approach was updated and extended, and a range of ontological problems resolved by Freeman (1991), whose account of the macrostructure of argument underpins most computational models of argument (Reed et al., 2007).

These different approaches vary in the types of nodes they allow, and in the types of informational connections between such nodes. They vary also in the extent to which they are merely informal tools for graphing arguments or are components of a wider system with a well developed syntax and semantics that affords computation over those graph structures. Examples of the latter include Dung's argumentation framework that represents arguments by means of directed graphs in which nodes represent individual arguments and directed links represent attack relations (Dung 1995). This framework has been used widely within artificial intelligence to model nonmonotonic reasoning. Another example is given by so called Bayesian belief networks which use directed acyclic graphs to represent variables and Independence relations between variables in order to simplify Bayesian computations. This framework has not just been used for probabilistic reasoning, but also for automated argument generation (Zuckerman et al., 1998), argument evaluation (Hahn, 2007), and machine generated explanation (Biris et al. 1999).

### *2.2. Technology for Argument mapping*

A plethora of tools to support argument mapping exist. In keeping with the wide variety of frameworks just discussed, these range from technology for visualisation through to systems that support varying degrees of computation over the represented graphs. There are now many reviews of such such systems – see Harrell (2005), Scheuer et al. (2010) and Kirschner et al. (2003) for prominent examples – but the classes of tools can be sketched with a small number of examples that can serve to demonstrate the strengths and weaknesses of the classes in general.

Rationale is a good example of a tool that focuses on argument mapping, laying out the structure of reasoning (van Gelder, 2007). It focuses on the connections between single ideas (often single sentences), distinguishing independent from co-dependent reasons, forming a tree structure. It provides a straightforward and accessible local-scale picture, but does not scale well to larger debates.

Debategraph (debategraph.net) aims to tackle this problem of scale directly by providing a 'windowed' view of a debate, centered on a particular idea with a limited horizon of connected information. Unlike many such tools, debategraph attracted a significant audience due in part to its use in several episodes of CNN's *Amanpour* televised news. From a technical perspective, it also overcame the 'tree' structure imposed by tools such as Rationale, instead allowing the construction of much more expressive graphs of interconnections. Debategraph though, like Rationale, uses ad hoc and idiosyncratic data representation methods that make it difficult to reuse and re-present the data.

In tools that have aimed to integrate argument processing rather than just argument representation, there have been examples that have tried to avoid this pitfall. Carneades (Gordon et al., 2007), for example, provides evaluative algorithms that are comparable to Dung (1995), but are founded in the work of Canadian philosopher Douglas Walton (Walton, 1996). Though technically very accomplished, such tools have been confined to the research lab and not scaled up.

Success at scale is very difficult in this area, and few systems have reached audiences even in the thousands. One notable exception is Kialo (kialo.com) which provides a platform for the structured expression of disagreement. Though kialo is designed for scale, and therefore has to be not only robust but also both easy and appealing to use, it has simplified its notion of argument structure so much that there is very little flexibility left. As a commercial entity, its data and platform are also closed, making wide-scale application at the science-policy interface more challenging.

For this work, we have adopted the Online Visualisation of Argument tool, OVA (Janier et al., 2014), as it balances several of the key concerns. First it is academic, open source, and supported. Second, it has grown out of research in argumentation theory and AI, and as a result, has a great deal of sophistication and nuance in its conception of argumentation, which allows significant flexibility in how debates are construed. Third, its data representation is the open standard, Argument Interchange Format, which has been stable now for over a decade (Chesñevar et al., 2006). And finally, it has scaled well, with over 100,000 users in 80 countries, and -crucially for a Collective Intelligence project- supports concurrent collaborative working. As a result of taking these facets together, it is also a part of an ecosystem of tools for data storage, manipulation and visualisation that can be tapped into (Reed et al., 2017).

### *2.3. Theoretical Issues*

It should already be clear from the preceding sections that there is no one true way to represent an argument. Rather, different frameworks and tools come with their own theoretical assumptions and perspectives which will be imposed on argumentative discourse. The selection of framework in which to pursue representation of scientific debate can consequently not be entirely theoretically neutral.

It may seem desirable to take a more bottom up approach and start from the actual scientific research itself and try to develop representation schemes that best suit scientific practice. However, at root of the difficulties lies the fact that text containing even only moderately complex argument will typically not have a unique, underlying structure in the first place. Rather, different analysts may see different overall structure in the same piece. It just makes little sense to construe the task as one of bringing out the true underlying structure; instead what representations are developed should be assessed on the basis of their adequacy for the task at hand.

In this context, it is also important to remember that our goal is not to map out the entire argumentative/rhetorical structure of individual research articles, but rather to document the evidence base for particular scientific *claims*. Individual journal articles may contain large passages, such as introductory reviews of past research or motivation of the current research question, that are partially, or even wholly, irrelevant to the actual evidence the article is adding to the evidence base.

Consequently, there seems little alternative to simply picking a particular framework and seeking to apply it in practice. Only this process will ultimately allow one to understand what works and what doesn't and what is ultimately required. We thus embarked on a pilot project which we next report as an initial case study.

## 3. A Case Study: Communicating Uncertainty

### *3.1. The Debate*

Given that we started this project motivated by the pandemic it made sense to choose a (then) policy relevant question within our research expertise. A fundamental debate within the literature on judgement and decision making concerns people’s ability to interpret different format for representing probabilities, as one might use in the communication of risk. In particular, there have been almost 3 decades of debate over the question of whether people find it easier to understand and reason with frequency information as opposed to probabilities. The theoretical debate we chose for our pilot project was thus: Are natural frequencies the default way people process uncertain information?

Communicating uncertainty is essential to practical policy - for example, it is imperative that governments and scientists communicate risks related to COVID-19 in ways where citizens best understand the risks and are able to make reasonable inferences concerning this information. However, it is unclear what format (if any) is best at communicating risk (for a meta-review, see McDowell & Jacobs, 2017). Uncertain information can be conveyed in different ways - it can be expressed as probabilities (0.4), it can be expressed as percentages (40%), and it can be expressed as natural frequencies (e.g. 40 out of 100). Within the psychological literature, researchers debate the effect (including if there is any) of different forms of communicating uncertainty.

Aside from being theoretically interesting, this question has direct application questions, as this debate influences how we may communicate uncertainty to the general public. If the format of the information influences the degree to which people make reasonable inferences, it is an important tool to better communicate risk and reward for issues such as benefits and risks related with vaccination, the probability of catching COVID-19, and consequences of catching COVID-19 with and without vaccination. In this example, comparative mortality with and without a vaccine can be communicated as probabilities, percentages, or natural frequencies. Finding the best way to communicate this information is key to clear and successful communication to citizens.

### *3.2. The Process*

While we ultimately envision the necessary steps from selection of relevant literature through to construction of the final argument map to be conducted as a collective intelligence project by a sizable group which could accomplish the tasks at speed, we thought it important to step through all aspects ourselves. Three of the authors of this paper (Author 1, 2, and 3) thus selected papers, identified arguments, and entered them in the mapping software in close collaboration with the fourth (Author 4), who was responsible for the software implementation.

Given the limited resources in terms of personnel and time, it would have been impossible to map out the entire relevant literature which consists of numerous articles across dozens of journals to represent competing viewpoints on the debate. In our step one, **identify relevant literature,** we selected four sample papers.

The first, a paper by Gigerenzer and Hoffrage (1995) (henceforth G&H) is the seminal publication on this issue. As such, we could not adequately represent the debate on communicating uncertainty without including this paper. The others are sceptical of the G&H position - however, for our purposes of illustrating academic debates via argument maps, they perform different and essential functions. First, Brase and Hill (2018) (B&H) provide a novel experimental study with evidence relating to assumptions in the G&H paper. This provides a way to illustrate disagreement on assumptions. Second, Girotto and Gonzales (2001) (G&G) primarily review other work to make a conceptual point. This allows us to represent disagreement on theoretical links rather than empirical assumptions. Finally, McDowell and Jacobs (2017) (M&J) provides a meta-review of the literature. This opens entirely different difficulties in representing how a debate is characterized in general and how to capture this in an argument map.

As such, the four papers represent a key study, an empirical disagreement, a theoretical disagreement, and a review of the debate. We considered these papers to be at least somewhat representative of the debate itself and, hopefully somewhat, representative of the kinds of argumentative contributions one might expect within this domain.

These papers in hand, we then proceeded to the second step, **mark up individual arguments**. It is fairly straightforward to identify the main claims in the sense of overall conclusions that are being advanced in each of the four articles.

Far more difficult is identifying the structure of the arguments/evidence supporting those conclusions in each of the papers. As indicated above, past experience suggests that different analysts may well come up with different reconstructions of the same argumentative text. For our first paper, G&H, each analyst thus started with their own independent analysis of the text. We then met to discuss these, identify issues and differences raised, and tried to come to an agreement.

Several key issues emerged. First is the level of detail to be encoded. To give a simple example, a paper may offer an experimental conclusion as an argument for the key claim at issue. That experimental conclusion is itself supported by the methods, the experimental design, the data, the statistical analysis, and the interpretation of the results. There is no in principle boundary for inclusion to be drawn here. Rather, decisions should be made with utility and transparency to the ultimate end user in mind. Furthermore, the more detail encoded, the more resource intensive the encoding process itself becomes.

Several considerations seemed important to us in resolving these issues. For one, as long as the reference to parts of text is clear, it should always be possible to go back to the original manuscript in search of yet more detail. Furthermore, detail will naturally come to be added if and when issues are contested. It is not the function of the mapping process we are conducting to itself critically analyse the strength of arguments or evidence. Rather, the goal is to communicate as evenhandedly as possible that debate.

It is a common presumption of dialectical exchange that positions not explicitly challenged are shared. Unless methodological details of a study themselves become issues within the scientific debate, it may be reasonable, in first instance, to repress them, or give only the briefest summary indication.

A further tool that should be available here is an ability to expand or collapse parts of the map, hiding or exposing detail in aid of comprehension. Such a technical feature becomes more and more essential as the overall argument map grows in size.

In light of this, what we converged on as a level of granularity is a mere first pass suggestion, and future research that included feedback from both scientific community and stakeholders would be required in order to determine suitable conventions on detail.

Finally, we felt it was important at all times to stick with the original text, rather than paraphrase. This is because paraphrasing necessarily introduces a further degree of analyst interpretation. It should be clear from the preceding discussion that a wholly objective rendition in which the analyst has no shaping rule is impossible. That does not alter the fact, however, that it is desirable to minimise subjectivity and interpretation as much as possible. The only tool we decided to allow with respect to length and detail was ellipsis, represented clearly with "…".

Decision on which material and how much detail to include is the first, and most fundamental, choice. The second is the question of how to express the relations between the different components of the overall argument. This will be determined to a large extent by the ontology that the mapping framework makes available. Given a choice between fine grained classifications of types of evidential support, and generic indications of support or conflict, our preference is strongly with the generic (e.g., "supports", or "default inference")–again, because it minimises the role of analyst subjectivity.

In principle, the reading, interpreting, and parsing a text into arguments and their interrelations could be conducted separately from the next step in the process, **transfer to machine readable template.** In practice**,** however, we found it convenient to conduct both steps together, marking up, selecting, and adding text to the map within the software. The final map we generated for the seminal G&H article is shown in Figure 1 below.

OVA allowed us to open a PDF from a repository we had set up, highlight text within that PDF, and transfer it to the emerging argument map, where links between individual arguments are then added. While we found available types of links within OVA, such as "default inference" indicating a generic support relationship, or "is a" indicating a definitional expansion, well suited to working with scientific text, we also ran into challenges.

In particular, citation as a ubiquitous feature of scientific text posed novel issues for the software. Multiple types of reference to other works recur regularly throughout scientific articles:

1. Supporting references for a claim made by the current author:… here is a scientific claim (see also, Smith 2000)
2. Paraphrases of others' claims: Smith and colleagues argued that bananas are yellow and bendy (Smith 2000)
3. Collections of authors:… yellow bananas (Smith 2000, Jones 2012, Miller 2020)

Conceptually, 1 and 3 point to a source that broadly will provide further evidence to support that claim. By contrast, for 2, the given reference *is* the claim, in the sense that both paraphrase and reference are two ways of individuating the same object (although the reference additionally provides a source to support that this is indeed the case). Different ways to mark these are consequently required.

In all cases, however, one would also like the reference ("Smith 2000") to be an object that can itself be linked to. This is because it seems of fundamental practical importance that the overall argument map of a scientific debate can be built up incrementally, because only then can this process be turned into one for collective intelligence. This required adjustment to the system.

One of the key advantages of a system such as OVA is that it largely automates the fourth step of the process, **build argument map.** This includes joining up sub-maps that share a common node. OVA straightforwardly does this because it simply allows users to import multiple individual maps (simply by opening multiple json files) and join them through common nodes.

It was thus possible for us to then individually each take one of the remaining papers, analyse these, and add them to the final overall map, as is essential for a collective intelligence process (see for an excerpt, Fig. 1).

### *3.3. Insights and Challenges*

Overall, our test case was encouraging and confirmed the feasibility of such an approach with extent technology. The technical challenges we encountered (as illustrated by the citation issue) tended to reflect both particularities of scientific discourse, and were solvable with minor adjustments or additions to the software. We expect that further work would identify additional challenges of this level, suggesting that, at present, such a project still requires the involvement of a software developer with access to the underlying platform. We are, however, optimistic that limited amounts of additional experience would allow us to progress in progress the approach to a point where further tweaks of the system would no longer be necessary, thus enabling large scale adoption of this approach.

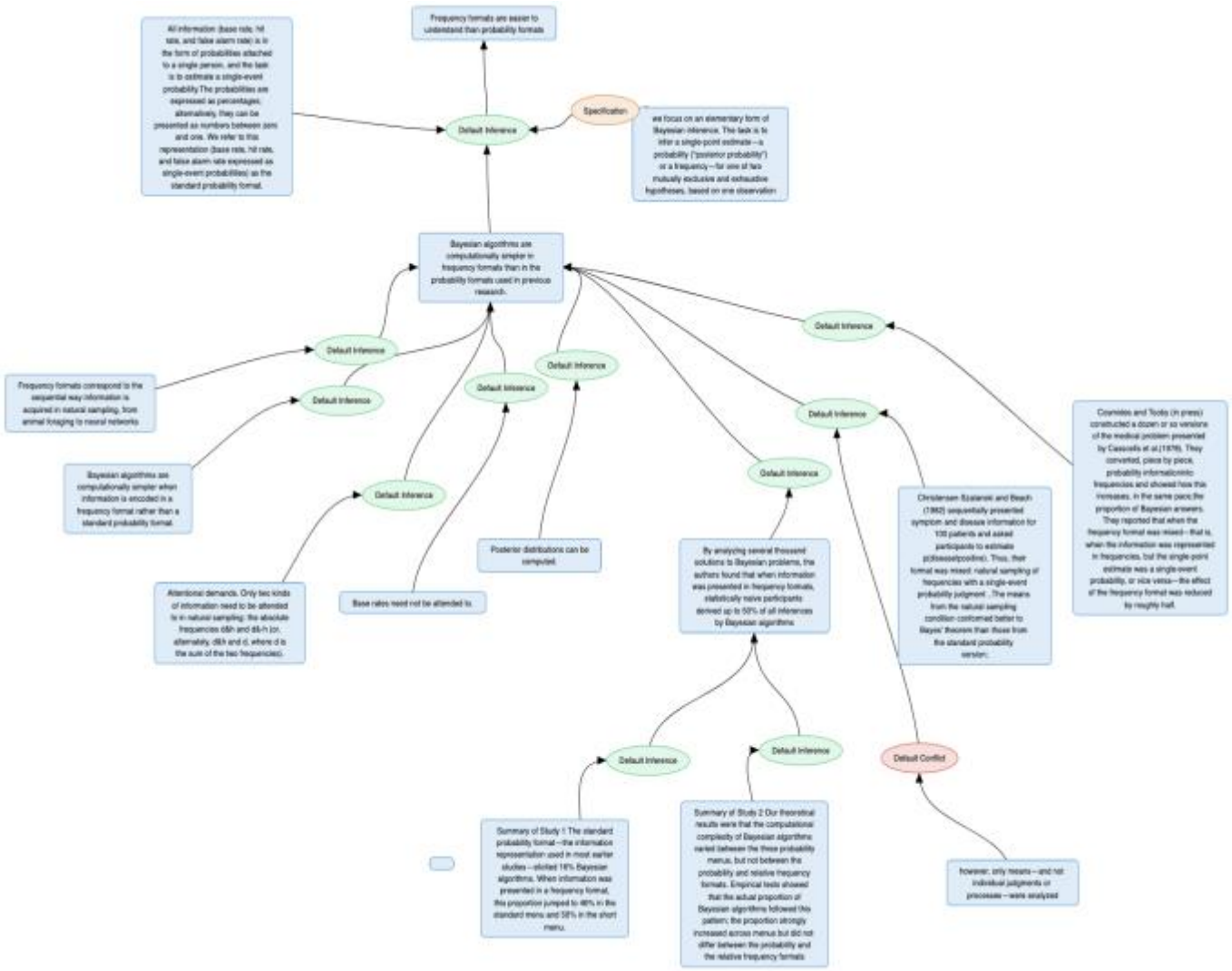

**Fig. 1 Argument map of Gigerenzer & Hoffrage  (1995)**

## 4. Alternative Means of Aggregation

In concluding, it seems useful to contrast the suggested argument map based approach with other conventional ways of formally summarising extent research. In particular, the suggested approach can be contrasted with three other ways in which researchers commonly synthesise research: critical reviews, systematic reviews, and meta analysis.

In a critical (narrative) review, a researcher or group of researchers presents their own overview and evaluation of a body of research. Crucially, while such reviews seek to provide a representative overview of extent research, it is entirely appropriate, and common, that material is presented and synthesised on the basis of the researchers own critical evaluation of the body of research. Presentation is thus shaped by what the authors themselves ultimately find theoretically compelling. Critical reviews are thus as much constructive as they are reconstructive with respect to the debate. This means also that evidence the author views as weak may be given only highly summarised mention or excluded altogether. Finally, given that aggregation is a key goal of critical reviews, much of the individual arguments or evidence will not be directly accessible within the review itself, and can be recovered only through use of the supporting references.

Systematic reviews differ from critical reviews in that they seek to make more objective the process of determining and reporting material to be included in order to minimise bias. In particular, systematic reviews require authors to identify selection criteria in advance of considering the research itself. Moreover, the selection process and the criteria used constitutes are described in the review itself. There is also a meta-science literature seeking to develop and articulate the process by which systematic reviews should be conducted (see, e.g., Meade et al., 1997). Systematic reviews are popular in disciplines such as medicine, where they may take on even more regimented forms, such as Cochrane reviews (see e.g., Henderson et al. 2010) -which follow detailed guidelines and are hosted on a dedicated platform (cochrane.org).

Meta analysis, shares with systematic reviews the fact that research selection for inclusion is formalised and typically reported within the article. The ultimate unit of analysis and the means of aggregation, however, are different. Specifically, meta analysis seeks to estimate the true effect size of an empirical phenomenon by statistically aggregating inferential statistics from multiple empirical studies (e.g., Lipsey & Wilson, 2001). Even more so than for systematic reviews, there is an extensive scientific literature on how to conduct meta analysis, and meta analysis may be deployed within systematic reviews.

Crucially, the desired statistical aggregation requires that the empirical studies are all fundamentally of the same kind. It would be conceptually meaningless to statistically aggregate experiments with very different design, let alone testing different consequences of a theory.

Meta analysis can thus only be applied to an otherwise very homogenous body of evidence. The systematic review allows somewhat greater flexibility here. However, even it does not fit naturally with very diverse types of supporting evidence such as those seen in the debate examined in this paper, and as is typical of the behavioural sciences. Our focal debate on the representation of probabilities includes experiments with different experimental designs, tests of different implications flowing from the fundamental theoretical positions, and it mixes empirical evidence with conceptual arrangements drawn from a wide variety of sources, all of which may relate in complex ways to one another and to the ultimate target claim or hypothesis.

An argument map seems uniquely placed to handle this diversity, and to draw out those multiple connections, while being more transparent, detailed, and balanced then a critical review. Clearly, not all end users and stakeholders may wish to engage to the same extent with that body of evidence. It is for this reason that we propose the aggregation steps not implemented in our pilot study. Crucially, unlike a critical review, the community-based voting we propose is both far more representative then an individual author opinion, as governs a critical review, and it also does not shape (and hence obscure) the evidence base itself. Finally, unlike a critical review, the full process we outline here in this paper is (like systematic reviews) amenable to implementation as a collective intelligence process that harnesses all available resources within a scientific community.

## 5. Summary

In this paper, we presented a novel alternative to the standard process for dispensing expert advice to policy makers. This process is clearly labour intensive, but, unlike traditional policy advice, does not suffer from extreme resource bottlenecks. Rather it is conceived as a process that draws on the resources of an entire scientific community, and in so doing affords transparency and balance that are presently frequently missing in high stakes policy contexts such as the pandemic. Our initial case study supports the feasibility of this idea, though further research would be required to converge on a best practice for all parts of this process. Likewise, further attempts to map out scientific debates are likely to surface issues we have not yet encountered. These issues, however, should be of interest both from a technical and an epistemological perspective. We thus hope to have spurred future interest in such work.

## References

[1] Biris, E., & Shen, Q. Automatic modelling using Bayesian networks for explanation generation. In *Proceedings of the 13th International Workshop on Qualitative Reasoning about Physical Systems* 1999 (pp. 19-26).

[2] Brase, G. L., & Hill, W. T. Adding up to good Bayesian reasoning: Problem format manipulations and individual skill differences. *Journal of Experimental Psychology: General*, 2018 *146*(4), 577.

[3] Chesnevar,C.,McGinnis,J.,Modgil,S.Rahwan,I.,Reed,C.,Simari,G.,South,M.,Vreeswijk,G.,Willmott,S. "Towards an Argument Interchange Format", Knowledge Engineering Review, 2006, 21 (4), pp293-316.

[4] Freeman, J.B. Dialectics and the macrostructure of argument, Foris. 1991

[5] Gigerenzer, G. & Hoffrage, U. it'sHow to improve Bayesian reasoning without instruction: frequency formats, *Psychological review* 1995. 102 (4), 684-704

[6] Girotto, V. & Gonzales, M. Chances and frequencies in probabilistic reasoning: rejoinder to Hoffrage, Gigerenzer, Krauss, and Martignon, *Cognition* 2001. 84 (3), 353-359

[7] Gordon, Thomas & Prakken, Henry & Walton, Douglas. The Carneades Model of Argument and Burden of Proof. Artificial Intelligence. 2007 171. 875-896.

[8] Hahn, U. Collectives and epistemic rationality. *Topics in Cognitive Science*.2 022

[9] Hahn, U., & Oaksford, M. The rationality of informal argumentation: a Bayesian approach to reasoning fallacies. *Psychological review*, 2007. *114*(3), 704.

[10] Harrell, M. Using argument diagramming software in the classroom. Teaching Philosophy, 2005, 5(28),163–177.

[11] Henderson, L. K., Craig, J. C., Willis, N. S., Tovey, D., & Webster, A. C. How to write a Cochrane systematic review. Nephrology, 2010. 15(6), 617-624.

[12] Janier, M., Lawrence, J., Reed, C. "OVA+: an Argument Analysis Interface" in Parsons, S., Oren, N., Reed, C. & Cerutti, F. (eds) Proceedings of the Fifth International Conference on Computational Models of Argument (COMMA 2014), IOS Press, Pitlochry, pp. 463-464.

[13] Kirschner, P., Buckingham Shum, S., & Carr, C. Visualizing argumentation: software tools for collaborative and educational sense-making. Springer. 2003

[14] Lewandowsky, S., Cook, J., Ecker, U. K., Lewandowsky, S., Cook, J., Ecker, U. K. H., & Newman, E. J. Under the Hood of The Debunking Handbook 2020: A consensus-based handbook of recommendations for correcting or preventing misinformation. 2021

[15] Lipsey, M. W., & Wilson, D. B. Practical meta-analysis. SAGE publications, Inc. 2001

[16] McDowell, M., & Jacobs, P. Meta-analysis of the effect of natural frequencies on Bayesian reasoning. *Psychological bulletin*, 2017. *143*(12), 1273.

[17] Meade, M.O., & Richardson, W.S. Selecting and Appraising Studies for a Systematic Review. Ann Intern Med.; 1997. 127:531-537. doi:10.7326/0003-4819-127-7-199710010-00005

[18] Page, S.E. The Difference: How the Power of Diversity Creates Better Groups, Firms, Schools, and Societies. By Scott E. Page. Princeton: Princeton University Press, 2007.

[19] Reed, C., Budzynska, K., Duthie, R., Janier, M., Konat, B., Lawrence, J., Pease, A. & Snaith, M. "The Argument Web: an Online Ecosystem of Tools, Systems and Services for Argumentation", Philosophy & Technology, 2017. 30 (2), pp137-160.

[20] Reed, C., Walton, D. & Macagno, F. "Argument diagramming in logic, law and artificial intelligence", Knowledge Engineering Review, 2007. 22 (1), pp. 87-109.

[21] Scheuer, O., Loll, F., Pinkwart, N., & McLaren, B.. Computer-supported argumentation: a review of the state of the art. International Journal of Computer-Supported Collaborative Learning, 5, 2010. 43–102.

[22] Suran S, Pattanaik V, Draheim D. Frameworks for collective intelligence: A systematic literature review. ACM Computing Surveys (CSUR). 2020 Feb 5;53(1):1-36.

[23] Toulmin SE. The uses of argument. Cambridge university press; 2003 Jul 7.

[24] van Gelder, T. The rationale for rationale. Law, Probability and Risk, 2007. 6(1–4), 23–42.

[25] Walton, D.N. XArgumentation Schemes for Presumptive Reasoning, LEA. 1996.

[26] Wigmore JH, editor. The principles of judicial proof: as given by logic, psychology, and general experience, and illustrated in judicial trials. Little, Brown,; 1913.

[27] Zukerman, I., McConachy, R., & Korb, K. B. Bayesian reasoning in an abductive mechanism for argument generation and analysis. In *AAAI/IAAI* 1998. (pp. 833-838).